\documentstyle[seceq]{ptptex}

\def\a'{\alpha'}
\def\s{\sigma}
\def\t{\tau}
\def\tension{\frac{1}{4\pi\alpha'}}
\def\sumk{\sum^{\infty}_{k=1}}

\def\calD{{\cal D}}
\def\p{{\partial}}

\def\mb#1{\mbox{\scriptsize{#1}}}
\def\mbt#1{\mbox{\tiny{#1}}}
\newcommand{\bea}{\begin{eqnarray}}
\newcommand{\eea}{\end{eqnarray}}

\notypesetlogo  

\title{
Closed-String Tachyon Condensation\\
 and\\ 
the On-Shell Effective Action of Open-String Tachyons
}

\author{
Shin N{\sc akamura}\footnote
{E-mail : nakashin@yukawa.kyoto-u.ac.jp}
}

\inst{
Yukawa Institute for Theoretical Physics, 
Kyoto University, Kyoto 606-8502, Japan}

\recdate{
May 26, 2001
}

\abst{
We study how the effect of closed-string tachyon condensation
can enter into the on-shell effective action of open-string 
tachyons in the bosonic case.
We also consider open-string one-loop quantum corrections
to the on-shell action.
We use a sigma-model approach with boundary terms, and
we utilize some results 
of boundary string field theory (BSFT)
to define the on-shell effective action.
We regard D-instanton-like objects with appropriate weight
as closed-string tachyon tadpoles, and we insert them into 
worldsheets to analyze the effect of closed-string 
tachyons.

}

\begin{document}

\maketitle

\section{Introduction}

One of the important subjects in string theory is the construction
of an off-shell formulation. Recently, studies of this subject
have proceeded in the context of open-string tachyon condensation,
using string field theories. \cite{CSFT,BSFT1,BSFT2}
String field theories provide tools to analyze the spacetime
action of open-string tachyons, and they have been utilized
\cite{SenZwi,Tay,MoeTay,HarKra,dMKJMT,MSZ,dMKR,Moe} 
to check Sen's conjecture. \cite{Sen1,Sen2,Sen3,Sen}
In particular, Sen's conjecture has been confirmed \cite{GerSha,KMM1,GhoSen} 
in the framework of boundary string field theory (BSFT).
\cite{BSFT1,BSFT2,BSFT3,BSFT4,BSFT5}
BSFT gives the exact spacetime action of open-string tachyons
at the tree level, \cite{GerSha,KMM1,GhoSen} and we can derive the
descent relation of such spacetime actions
on D-branes of various dimensions. \cite{MoriNaka}
Currently, we know that an unstable D-brane decays completely
when the open-string tachyons
living on it fully condense.
Therefore, it is thought that the open-string sector 
disappears after the open-string tachyon condensation, 
and only the closed-string sector is left. 

Although studies of open-string tachyons have progressed, 
it is still difficult to treat closed-string tachyon
condensation. 
No one knows precisely what happens after closed-string 
tachyon condensation.
Therefore, determining the fate of both the open-string sector 
and the closed-string sector is very interesting, 
and studies of closed-string 
tachyons represent an important approach to making 
such a determination. 

In this paper, 
we attempt to treat {\em closed}-string tachyon condensation
in bosonic strings.
However, we do not consider off-shell closed-string dynamics.
Our aim here is to calculate physical quantities
on the background in which closed-string tachyons have
already condensed.
Specifically, 
we consider the {\em on-shell} effective action of
bosonic open-string tachyons on the background in which 
closed-string tachyons have already condensed.
We also consider open-string one-loop quantum corrections
to such an on-shell effective action.
In this article,
we do {\em not} attempt to obtain {\em off-shell} effective 
action of open-string tachyons on the background in which 
closed-string tachyons are condensed. 
It is quite difficult to construct open-string field
theories on such a background, although those on 
the usual flat background without closed-string
tachyon condensation have been constructed.
With these points in mind, this paper is intended 
as an investigation of a method of obtaining a consistent
{\em on-shell} effective action of open-string tachyons
{\em on the background} in which closed-string tachyons have
condensed.
We propose a basic procedure to obtain a consistent
on-shell effective action.
Basically, we use a sigma-model approach. \cite{FeadTsyt85,Tsyt88}
We also utilize the results of BSFT to define the on-shell 
effective action of open-string tachyons for each vacuum.
We treat oriented strings for simplicity.

The organization of this article is as follows.
In \S 2, we present a basic strategy to treat closed-string
tachyon condensation. The method we use here is based on the
model that was originally proposed by Green. 
\cite{Green76,Green77,GreenShap76,Green88,Green91}
We regard D-instanton-like objects with appropriate weights
as closed-string tachyon tadpoles, and we insert them into 
worldsheets to analyze the effect of closed-string 
tachyons.
In \S 3,
we propose a sigma-model approach with boundary terms
to analyze the corrections 
from closed-string tachyons and
open-string loops.
We point out that we can utilize the results of
BSFT to define the on-shell effective action of open-string
tachyons.
Although it is very difficult to estimate the effect of
closed-string tachyon condensation precisely,
we calculate the modified 
on-shell effective action of open-string
tachyons for some special cases in \S 4 and in 5.
In general, the
on-shell effective action of open strings possesses an instability 
due to closed-string tachyons
if loop corrections are included.
We propose a basic procedure to obtain a consistent
on-shell effective action that has no instability
due to closed-string tachyons in \S 5.
Although it is still difficult to fix the weight of the
closed-string tachyon tadpole, the condition that
the instability due to closed-string tachyons vanishes 
provides a constraint on the weight.
Using the procedure we propose there, we can obtain 
a finite correction to the on-shell effective action,
at least at the one-loop level of open strings,
in principle.
In the final section, we give a conclusion and make some remarks.
A problem involving the equation of motion in BSFT is
pointed out in the Appendix.
We also propose a method of avoiding this problem
in order to utilize the results of BSFT.

\section{Tachyon condensation and tadpoles on the worldsheet}

In this section, we recall a previous attempt to describe strings
on the background in which tachyons are condensed.
First, let us recall the formulation of field condensation 
in usual quantum
field theories. In point-particle field theory 
with field condensation (that is, 
in the theory with a nonzero expectation value of the field), we can
calculate correct quantities if we know the correct expectation value 
of the field, even with a perturbation around an incorrect vacuum.
For example, we can calculate the exact propagator with a perturbation
around a tachyonic vacuum by attaching tadpole diagrams to the tree
propagator.\footnote{
See for example Ref.~\citen{coleman}.
}
Even though the mass squared is negative in a description 
around such a vacuum, tadpoles with appropriate weight create 
an additional shift of the mass squared, and cause the total 
mass squared to become positive (or zero).
The weight attached to the tadpole corresponds to the expectation
value of the field.
In such a case, although the tachyonic mode exists in a perturbative 
theory around a tachyonic vacuum, the theory is not incorrect, and only
the ``vacuum'' is incorrect. In field theories, the true vacuum or 
exact expectation value of the field can be obtained using the
Schwinger-Dyson equation. In this way we can get the correct weight of 
the tadpoles, we can calculate the correct propagator, and so on.

The important point here is that we do not need the exact Feynman rules
around the true vacuum.
We can reproduce them with the Feynman rules around the
tachyonic vacuum if we attach tadpoles to the graphs
with appropriate weight.
Therefore, this method is suitable even for first
quantized theories.

We now draw an analogy between field theories and 
string theories and introduce tachyon ``tadpoles'' into the 
framework of string theory.\footnote{
Attempts to describe tachyon condensation
using the emission of massless scalar particles from
string diagrams are found in Refs.~\citen
{Bard74}--~\citen{Bard78}
.
}
Green proposed many years ago that an off-shell extension
of the string amplitude is realized by insertion of Dirichlet 
boundaries into a worldsheet. \cite{Green76,Green77,Green88,Green91} 
We naturally assume that the inserted Dirichlet 
boundaries correspond to the tadpoles in string theories.
We also assume that
macroscopic holes with Dirichlet boundary conditions on the 
worldsheet represent closed-string tachyon tadpoles and
Dirichlet boundaries inserted into the boundary of the worldsheet
represent open-string tachyon tadpoles.
The Dirichlet boundary conditions we use are ``D-instanton-like''
boundary conditions, namely,
\bea
X^{\mu}(\xi_{i})={x_{0}^{\mu}}_{(i)},
\label{tad-1}
\eea
where the subscript $i$ distinguishes the tadpoles, 
and $\xi_{i}$ denotes the 
worldsheet coordinate where the tadpole is inserted.
${x_{0}^{\mu}}_{(i)}$ is a constant
that corresponds to the position of the tadpole in spacetime.
Therefore, each Dirichlet boundary is mapped to a single point 
${x_{0}^{\mu}}_{(i)}$ in spacetime.
The contribution of the D-instanton-like tadpole 
is integrated over ${x_{0}^{\mu}}_{(i)}$
with appropriate weight.
Therefore, if the weight does not depend on the spacetime
coordinates, translational invariance is restored.
It is known that these point-like energy densities on strings 
greatly alter the behavior of the strings. \cite{Green91}

The reasons we regard the above D-instanton-like 
Dirichlet boundaries as tachyon tadpoles are 
as follows. \cite{Kawai},\footnote
{ Most of the following arguments
already appear in Ref.~\citen{Nakamura}, 
in which an attempt is made to construct consistent noncritical strings.},\footnote
{Some argument for the macroscopic hole as a tachyonic state
 is also given in Ref.~\citen{Seib90}
.}

\vspace{0.5cm}

\begin{itemize}

\item Tachyon tadpoles are off-shell in general.
For example, if the expectation value of a tachyon does
not depend on the spacetime coordinates, the weight should
be constant and tadpoles do not carry momentum.
In such a case, tachyon tadpoles are off-shell because
tachyons have non-zero mass squared. 
We also know that off-shell
states in string theories do not correspond to local emission 
vertices. Thus we naturally assume that a tachyon tadpole 
is a non-local macroscopic boundary inserted into the worldsheet.
Namely, the tadpole for a
closed-string tachyon should be a macroscopic hole in the 
worldsheet, and the open-string tachyon tadpole should be
a macroscopic line inserted into the boundary of the
worldsheet.

\vspace{0.2cm}

\item We assume that we have a stable bosonic-string vacuum,
which might not yet be known.
We also assume that we have some dynamics that lead us
there from a $26$-dimensional flat tachyonic vacuum.
Our goal here is to obtain the on-shell quantities 
of strings around such a stable vacuum, in the case that it exists.
More precisely, we attempt to represent physics just
{\em on the stable vacuum} using well-known 
string Feynman rules obtained around the tachyonic vacuum.
Therefore, we must preserve Weyl invariance on 
the worldsheet even if we include tachyon tadpoles.\footnote 
{If we wished to describe off-shell dynamics of strings, 
we should insert macroscopic boundaries as tadpoles
that break Weyl invariance.}
It is natural to impose Dirichlet boundary conditions on
the macroscopic boundaries as tadpoles, which maintain Weyl 
invariance. 
This is because boundaries with Neumann boundary conditions
can emit on-shell open strings,
and this does not seem to be a natural property for tadpoles
that are coupled to an external (tachyon) field.
In this sense, Neumann boundaries would correspond to
the tadpoles which represent vacuum polarization when
the emitted strings form loops.

\vspace{0.2cm}

\item It has been reported that the insertion of
D-instanton-like boundaries into worldsheets alters the
vacuum state of string theories radically. \cite{Green91}
In addition, the naive estimation in Ref.~\citen{Nakamura} 
suggests that they
can shift the mass squared of tachyons.

\vspace{0.5cm}

\end{itemize}

Although we do not have rigorous proof that
the macroscopic boundaries mentioned above
represent tachyon tadpoles in string theory,
we call these D-instanton-like Dirichlet boundaries 
``tachyon tadpoles'' in this paper;
a macroscopic hole with the above Dirichlet boundary condition is
a closed-string tachyon tadpole, and an inserted Dirichlet boundary
on the edge of the worldsheet is an open-string tachyon tadpole.
We emphasize that they are not D-instantons as solitonic solutions
of string theories.
D-instantons are physical objects with definite tension
(although they might be unstable), while tachyon tadpoles
are parts of Feynman diagrams with the weight attached to them
corresponding to the expectation value of the tachyon field.

We insert closed-string tachyon tadpoles into the worldsheet
in order to reproduce the background in which closed-string 
tachyons are condensed.
However, we do not use open-string tachyon 
tadpoles in this paper for simplicity, 
because we can utilize the results of BSFT
to describe open-string tachyon condensation.

Of course, we must consider the proper weights of the string wave
functions on the tadpoles.
Unfortunately, we do not have a Schwinger-Dyson equation for 
string theory, 
and we do not know how to obtain the correct weight.
Thus, we cannot give a rigorous treatment of closed-string
tachyon condensation. However, we derive a constraint for the
weight in \S 5.

\section{Sigma-model approach and closed-string tachyon
condensation}

In this section, we present a basic strategy to analyze
the on-shell effective action of open-string tachyons
on the background in which closed-string tachyons are condensed.
We basically use a sigma-model approach.
We also utilize the results of BSFT to define the partition
function of open strings and to identify the
vacuum of open-string tachyons.
Therefore, we begin with a brief review of 
the sigma-model approach.

\subsection{Sigma-model approach on the background in which
closed-string tachyons are condensed}

The basic idea of the sigma-model approach \cite{FeadTsyt85,Tsyt88} 
is that the
spacetime action $S$ for string fields
is provided by the renormalized partition function $Z$
in the background fields.
That is,
\bea
S=Z.
\label{sigma}
\eea
For example, the tree-level spacetime action for open strings
is given as the partition function derived on the disk worldsheet
with background fields on the boundary.
However, the relation (\ref{sigma}) is not correct
in the off-shell region, where we do not have conformal
symmetry on the worldsheet, for bosonic strings in general.
Therefore, we use (\ref{sigma}) only for the case
in which we calculate the on-shell spacetime action.

Next, let us extend the idea of the sigma-model approach to
the case that we consider open strings on the background
in which closed-string tachyons are condensed.
As discussed in \S 2, the worldsheet on the stable
vacuum can be reconstructed with the worldsheets
on the tachyonic vacuum.
More precisely, the effect of closed-string
tachyon condensation is represented by
the worldsheets to which closed-string tachyon
tadpoles are attached.
The partition functions obtained on such worldsheets
would correspond to the
correction terms to the spacetime action.
In this article,
we assume that the closed-string coupling $g_c$
is sufficiently small, and we calculate only annulus
diagrams as the correction terms.
In this case, we should also consider the annulus
worldsheet that represents an ordinary open-string
one-loop quantum correction, since it has the same
topology.\footnote{
We are not sure that this treatment is completely valid.
We discuss a related problem in \S 6.
}
The difference between the two types of annuli lies in the
boundary conditions imposed on the inner circle.
The worldsheets that represent the quantum corrections
have the same boundary conditions on the inner circle
as those on the outer circle,
while the diagrams that include closed-string tachyon 
tadpoles have inner circles on which D-instanton-like boundary
conditions are imposed.
We regard the outer circle as an ordinary boundary
of an open-string worldsheet that is coupled to a general
D-brane.

Therefore,
if we expand the total partition function with respect to
the closed-string coupling constant, 
we obtain
\bea
Z^{\mb{total}}&=&
Z_{0}
+g_{c}\int_{0}^{\infty} \frac{dt}{2t} 
\prod_i 
\int\!\! \frac{dx^{\mu}_0}{\sqrt{2\pi \a'}} 
w(x^{\mu}_0)Z_{1}^{\mb{tad}}(t,x^{\mu}_0) \nonumber \\
&+&
g_{c}\int_{0}^{\infty} \frac{dt}{2t} 
Z_{1}^{\mb{loop}}(t)
+O(g^{2}_{c}),
\label{total}
\eea
where $x^{\mu}_0$ denotes the position of D-instanton-like tadpoles
in the spacetime,
$w(x^{\mu}_0)$ is the weight attached to the 
tadpoles, which depends on $x^{\mu}_0$ in general, and
$t$ is the moduli of the annulus, 
which is defined in \S 4.

The leading order term $Z_{0}$ is the partition function
on the disk with no tadpole and no loop. 
The terms of order $g_{c}$ consist of two parts;
one of them contains the 
partition function $Z_{1}^{\mb{tad}}(t,x^{\mu}_0)$, which has
one closed-string tachyon tadpole, and the other is composed of
the partition function
$Z_{1}^{\mb{loop}}(t)$, which has one open-string loop.


In this paper, as an example of a concrete calculation
of the effect of closed-string tachyon condensation, 
we attempt to calculate the on-shell
effective action of open-string tachyons on the
background in which closed-string tachyons are condensed.
We write this effective action as $S_{\mb{cond}}$.
Basically, we assume the relation $S_{\mb{cond}}=Z^{\mb{total}}$
in the framework of the sigma-model approach.
In this sense, the sigma-model approach provides
a sufficiently powerful tool for us. 
However, as we find bellow, the boundary term
in the worldsheet action 
used in BSFT is useful to define the on-shell 
value of $Z^{\mb{total}}$ even in sigma-model approach.
For this reason, we make a brief review of BSFT in the
next subsection.

\subsection{A short review of BSFT for the bosonic case}

In the framework of the sigma-model approach, 
the dependence on the renormalization
scheme of the partition function is considered to 
be equivalent to the field redefinition of $S$.
This equivalence holds as long as the divergence
of the partition function is logarithmic.
However, it was pointed out that a power-law divergence
breaks this equivalence, and there results an ambiguity
in the manner of obtaining
$S$ from the partition function in such a case. 
\cite{AndTsey-1,AndTsey-2}
String theories that include tachyons have this 
problem. \cite{Tsyt88}

BSFT provides us a solution to this problem.
It gives us a natural extension
of the relation (\ref{sigma}) in the sigma-model approach.\footnote{
Therefore, what we learn that
BSFT can be applied to studies of the sigma-model
approach. Recent studies of this kind are also found in
Refs.~\citen{Tsey00}--~\citen{And01}
.
}
The relation between the spacetime action $S$ and the
partition function $Z$ in BSFT is given as
\bea
S(\lambda^i)=Z(\lambda^i)
\biggl(
1+
\sum_{i}\beta^i(\lambda^i)\frac{\p}{\p\lambda^i}\log Z(\lambda^i)
\biggr),
\label{sandz}
\eea
where $Z(\lambda^i)$ is the partition function given by
\bea
Z(\lambda^i)=
\int\calD X
 \exp\{-S_{w}(X,\lambda^i)
\},
\label{Z-lambda}
\eea
and the $\beta^{i}(\lambda^i)$ are $\beta$-functions obtained with
the worldsheet action $S_{w}$.
The worldsheet action for the construction of BSFT for 
open-string tachyons is
\bea
S_{w}
=\tension \int_{\Sigma}\! d\s d\t \partial_{a}X^{\mu}\partial^{a}X_{\mu}
+\tension \int_{\partial \Sigma}\!\!\!\! d\theta 
\bigl\{
2\a'a+\sum^{26}_{i=1}u_{i} (X(\theta)^{i})^{2}
\bigr\}.
\label{worldsheet}
\eea
The quantities $\lambda_i$ in (\ref{sandz}) and (\ref{Z-lambda})
are coupling constants included in the boundary
term of $S_{w}$;
they are $a$ and the $u_i$ in this case.
They control the configuration of the open-string tachyon
$T_{\mb{open}}$ through the relation
\bea
T_{\mb{open}}=a+\sum^{26}_{i=1} \frac{u_{i} {X_{i}}^{2}}{2\alpha'}.
\eea
The worldsheet $\Sigma$ is chosen to be a disk, on which
we have rigid rotational symmetry.
We do not need conformal symmetry in this case,
and the boundary term in (\ref{worldsheet}) breaks
conformal invariance, in general.

Then the BSFT action for open-string tachyons is given by
\bea
S(a,u_i)=Z(a,u_i)
\biggl(1+a+\sum_iu_i-\sum_iu_i\frac{\p}{\p u_i}\ln Z(a,u_i)\biggr),
\label{BSFT}
\eea
where $Z(a,u_i)$ is the partition function 
defined by (\ref{Z-lambda}) and (\ref{worldsheet}).
It is given exactly \cite{BSFT2} by
\bea
Z(a,u_i)=
A
e^{-a}
\Bigl(\prod_{i=1}^{26}\sqrt{u_i}e^{\gamma u_i}\Gamma(u_i)\Bigr)
\equiv
Z_0(a,u_i).
\label{Z-zero}
\eea
Here $A$ is a normalization factor, which is determined
so that $S(0,0)=Z(0,0)=V_{25}T_{25}$, where $V_{25}$ is the
volume of D25-brane and $T_{25}$ is the tension of
D25-brane.
We find that $A$ should be fixed as $A=T_{25}{(2\pi\alpha')}^{-13}$ 
in the Appendix.
The precise definitions of S(0,0) and Z(0,0) are given bellow.

If we do not consider a closed-string sector,
the tree level spacetime action of open-string tachyons is
given by (\ref{BSFT}).
The action (\ref{BSFT}) has several classical
solutions which correspond to unstable D-branes
of various dimensions.
If we substitute the corresponding classical solution
$(a_{\ast},u_{\ast i})$ into the action, $S(a_{\ast},u_{\ast i})$ 
gives the energy density times the volume of the 
corresponding D-brane.
Explicitly, we obtain the tension of the corresponding
D-brane if we divide the on-shell value of $S(a,u_i)$
by the volume of the spacetime where
open strings exist.
%
The value of $a_{\ast}$ and the $u_{\ast i}$ are $0$ or $\infty$.
We obtain the classical solution $(a_{\ast},u_{\ast i})$
by taking the limit discussed bellow.

Note that (\ref{sandz}) becomes the equivalent to
(\ref{sigma}) if we have conformal symmetry on the worldsheet.
Conformal symmetry is recovered
if we take the limit
\bea
a &\longrightarrow&
\left\{
\begin{array}{c}
0\\
\infty \\
\end{array}
\right. 
,\nonumber \\
u_i &\longrightarrow&
\left\{
\begin{array}{c}
0\\
\infty \\
\end{array}
\right.
.
\label{limit}
\eea
In general, we take a limit in which $n_{0}$
of the quantities $u_i$
go to zero and $n_{\infty}$ of the $u_i$ go to $\infty$,
where $n_{0}+n_{\infty}=26$.
If we take the above limit while imposing
the stationary condition for the variable $a$,
the variables $a$ and $u_i$ converge to a classical solution 
given by $a_{\ast}$ and $u_{\ast i})$.
The stationary condition we use here is
obtained in the Appendix as
\bea
a=
\sum^{26}_{i=1}
\left(
-u_i+ u_i\frac{\p}{\p u_i}\ln Z_0(u_i)
\right)
+
\sum_{s}
\left(
\frac{1}{2}-\frac{u_s x_s^2}{2\alpha'} 
\right),
\label{statcond}
\eea
where the values $u_s$ in the last term are those taken to zero
in the conformal limit, and therefore the constant term
coming from the sum in the last term is $\frac{n_0}{2}$.
The values $x^s$ are the spacetime coordinates parallel to the D-brane,
and they correspond
to the zero modes of the $X^{s}$.
We can check easily that $S(a,u_i)=Z_0(a,u_i)$ if
(\ref{statcond}) is satisfied.
Then, the on-shell action is obtained as
\bea 
S(a_{\ast},u_{\ast i})=
\lim_{
(a,u_i) \rightarrow (a_{\ast},u_{\ast i})
}
\!\!\!\!\!\!\!\!\!\!\!\!S(a,u_i)
=
\lim_{
(a,u_i) \rightarrow (a_{\ast},u_{\ast i})}
\!\!\!\!\!\!\!\!\!\!\!\!Z(a,u_i),
\eea
where the limits are taken along (\ref{statcond}).
The limit $u_i \rightarrow \infty$ yields Dirichlet boundary
conditions for $X^i$, and therefore $n_{\infty}$ corresponds to 
the number of directions
that are perpendicular to the corresponding D-brane.
Here, the action (\ref{BSFT}) becomes the tension
times the volume of 
D$(n_{0}-1)$-brane after we take the above limit. 
Actually, there are some delicate problems
concerning the method of taking 
the conformal limits.
We discuss the relevant details in the Appendix. 

In the original spirit of BSFT,
(\ref{BSFT}) is the spacetime action
that describes open-string tachyons even in the off-shell region.
BSFT might not be necessary for us, 
since we wish to obtain the on-shell
action, which can also be obtained from the sigma-model approach.
However, the boundary term that
includes the variables $a$ and $u_i$ also provides
us a good definition of the on-shell partition function
through the above conformal limit,
even within the sigma-model approach;
the $u_i$ play the role of a {\em regulator}, and the combination
of $a$ and the $u_i$ acts as a {\em navigator}, which leads
us to each vacuum through the above conformal limit.

The partition function on a disk without boundary terms
has an IR divergence.
The quantities $u_i$ act as regulators, which regularize
the IR divergence through integration over the zero-modes
of $X^{\mu}$, which correspond to the spacetime coordinates.
Positive $u_i$ prevent such divergence.
Furthermore,
we calculate the partition function with the boundary term in
(\ref{worldsheet}) and take the conformal limit (\ref{limit})
while imposing the relation (\ref{statcond}). 
Then we obtain the on-shell
effective action at each vacuum for open strings.
In this sense, the variables $a$ and $u_i$ play the role
of a navigator that leads us to each vacuum of the open-string
tachyons;
$n_0$ determines the dimension of the D-brane where the
open strings exist, and (\ref{limit}) with (\ref{statcond})
provides us the correct normalization of the partition
function at each vacuum.

It is hoped that we can construct an extended
BSFT that also describes off-shell dynamics of open-string tachyons
on the background in which closed-string tachyons are
condensed, using the partition function $Z^{\mb{total}}$
instead of $Z_0$.
We might also be able to include quantum corrections
in BSFT.\footnote{
Some attempts to obtain a loop-corrected BSFT action
are found in 
Refs.~\citen{RasVisYang}--~\citen{Cra-Kra-Lar}
.
}
However, there are some subtleties involved in the construction
of such an extended BSFT, as we discuss in the next subsection.
Therefore, we do {\em not} attempt to extend BSFT here,
and we carry out our analysis only for the on-shell properties 
of strings on the basis of the sigma-model approach,
including $O(g_c)$ corrections.
Thus, we do {\em not} regard the quantities $a$ and $u_i$ as dynamical
variables, as those in BSFT.
Instead, 
we use the results of BSFT only for the definition of the on-shell
effective action of open-string tachyons in the framework
of the sigma-model approach.

To restate our position, we are not concerned with 
off-shell actions in this paper.
In this case, we are not able to obtain the
profile of the open-string tachyon potential.
However, we can obtain the absolute value of the
effective action of
open-string tachyons at each vacuum
on the background in which closed-string tachyons are
condensed.

\subsection{Subtleties in the construction of BSFT 
including $O(g_{c})$ corrections}

In order to represent the dynamics of open-string tachyons
on the background in which closed-string tachyons are condensed, 
it may be possible to
construct BSFT using a partition function that includes
$O(g_{c})$ terms.
However, there are some nontrivial problems involved in the naive
extension of BSFT.
These subtleties are as follows.

First, in the definition of BSFT, \cite{BSFT1,BSFT2} we consider
a disk worldsheet with rotational symmetry.
Rigid rotational symmetry is required to obtain the
BSFT action.
However, it seems that we still have various choices 
of how to insert macroscopic
holes into the rotationally invariant disk
when we construct worldsheets that have an annulus
topology.
Most of the worldsheets we obtain after the insertion of
a hole do not have rotational symmetry.
In the on-shell case, we have Weyl invariance, and
we can map all of such worldsheets to an annulus that
is rotationally invariant. 
However, the boundary term that breaks Weyl
invariance does not allow us to naively
transform the worldsheet to an annulus.
Furthermore, we lose rotational symmetry if we insert
more than one hole, even in the case that we have Weyl symmetry.

Second, even if we have a good idea of how to construct BSFT
with worldsheets of arbitrary topology,
it is difficult to calculate contributions of all the worldsheets
of different configurations 
without conformal invariance.

\section{Correction terms at order $\boldmath{g_c}$}

\subsection{Setting up the analysis}

We now attempt to calculate the on-shell spacetime action of
open-string tachyons 
on the background in which closed-string tachyons 
are condensed.
Although the exact calculation of the effect of
the insertion of multiple tadpoles and open-string loops is very 
difficult, we can calculate the partition function
at order $g_c$.
According to the argument in the previous sections,
we must calculate $Z_1^{\mb{tad}}$
and $Z_1^{\mb{loop}}$.


To begin with, let us consider only a partition function with
one closed-string tadpole, in order to simplify the analysis.
An ordinary open-string one-loop partition function
can be obtained easily, and we mention it later.

First, we consider the annulus $M$. 
The inner circle of the annulus represents the 
tadpole, and we impose
D-instanton-like Dirichlet boundary conditions
on it.
The outer circle of the annulus is the usual
boundary of the worldsheet, which couples to the
open-string tachyons. We include the boundary term
of the worldsheet action only for the outer
circle.
We set the radius of the outer circle to $1$
and that of the inner circle to $r$.
This annulus becomes the disk that is used
to define BSFT if we remove the inner boundary.
The worldsheet action on the annulus $M$ is given by
\bea
S_{\mbox{\scriptsize worldsheet}}
&=&\tension \int_{M}\! d\s d\t \partial_{a}X^{\mu}\partial^{a}X_{\mu}
\nonumber \\
&+&\tension \int_{\partial M_{\mb{out}}}\!\!\!\! 
d\theta 
\bigl\{2\a' a+\sum_{i=1}^{26}u_{i} X_{i}(\theta)^{2}
\bigr\},
\label{annu-action}
\eea
and the boundary conditions for the $X^{\mu}$ are
\bea
n^{a}\partial_{a}X^{\mu}+u_{\mu} X^{\mu}{|}_{\partial M_{\mb{out}}}=0 ,
\label{cond-1}
\eea
\bea
X^{\mu}{|}_{\partial M_{\mb{in}}}=x_{0}^{\mu},
\label{cond-2}
\eea
where
$\partial M_{\mb{in}}$ and 
$\partial M_{\mb{out}}$ denote the inner boundary and
the outer boundary, respectively.
Here $n^{a}$ is the unit normal vector.
Note that we do not sum over $\mu$ in the second term of the
left-hand side of (\ref{cond-1}).

We also must integrate the partition function
over the moduli. We have two types of moduli here.
One of them is $r$, the radius of the inner circle of
the annulus.
The other is $x_0^{\mu}$, the position of the D-instanton-like
tadpole in spacetime.
After the integration over $x_0^{\mu}$, 
we take the conformal limit to justify
our calculation. Then we fix the gauge, and integrate
over $r$.
Although the weight $w(x^{\mu}_0)$ can depend on
$x^{\mu}_0$ nontrivially in general, 
we have no technique to determine it rigorously as yet.
Therefore, we regard $w$ as a constant in this section.
An argument regarding the $w(x^{\mu}_0)$ dependence of $w$ 
is made in \S 5.

\subsection{Calculation of $Z_1^{\mb{tad}}$}

Now we begin the calculation of $Z_1^{t\mb{ad}}$.
We have four types of parameters that control
the partition function. They are $a$, $u_i$, $x_{0}^{\mu}$
and $r$. We can divide the calculating procedure into
several parts as follows.

\subsubsection{The $u_i$ dependent part and the $a$ dependent part}

First, we calculate the $u_i$ dependent part
of the partition function with the action (\ref{annu-action})
employing the method in Refs.~\citen{BSFT2} and ~\citen{Suyama}.
Let us first consider the case in which only one of $u_{i}$ is non-zero
and the others are zero.
We omit the subscript $i$ in this case.

First, we set $z=\s+ i\t$.
The components of the metric on the annulus $M$ are 
$g_{z\bar{z}}=g_{\bar{z}z}=\frac{1}{2}$,
$g_{zz}=g_{\bar{z}\bar{z}}=0$.
The Green's function on the annulus should obey
\bea
-\frac{1}{\pi \a'}\partial_{z}\partial_{\bar{z}}G(z,w)=\delta^{2}(z,w),
\eea
and the boundary conditions are
\bea
\left( 
(z\partial_{z}+\bar{z}\partial_{\bar{z}})G(z,w)+u
\right){|}_{|z|=1}
=0,\\
G(z,w){|}_{|z|=r}=x_{0}^{\mu}.
\eea

To begin with, we calculate the partition function for $x_{0}^{\mu}=0$,
and then we determine the $x_{0}^{\mu}$ dependent factor.
If $x_{0}^{\mu}=0$, the above stated requirement determines 
the Green's function to be
\bea
G(z,w)=&-&\frac{\a'}{2}\ln|z-w|^{2}-\frac{\a'}{2}\ln|1-z\bar{w}|^{2}
-\frac{\a'}{4}\frac{u}{1-u\ln r}\ln|z|^{2}\ln|w|^{2} \nonumber \\
&+&\frac{\a'}{2}\frac{1}{1-u\ln r}\left(\ln|z|^{2}+\ln|w|^{2}\right)
-\a'\frac{\ln r}{1-u\ln r} \nonumber \\
&-&\frac{\a'}{2}\sumk
\left(\frac{1}{k}\frac{(k-u)r^{2k}}{k+u+(k-u)r^{2k}}\right)
\left\{
{\left(\frac{z}{w}\right)}^{k}+{\left(\frac{z}{w}\right)}^{-k}
+{\left(\frac{\bar{z}}{\bar{w}}\right)}^{k}
+{\left(\frac{\bar{z}}{\bar{w}}\right)}^{-k}
\right\} 
\nonumber \\
&-&\frac{\a'}{2}\sumk
\left(\frac{1}{k}\frac{2u+(k-u)r^{2k}}{k+u+(k-u)r^{2k}}\right)
\left\{{(z\bar{w})}^{k}+{(\bar{z}w)}^{k}\right\} \nonumber \\
&-&\frac{\a'}{2}\sumk
\left(\frac{1}{k}\frac{(k+u)r^{2k}}{k+u+(k-u)r^{2k}}\right)
\left\{{(z \bar{w})}^{-k}+{(\bar{z} w)}^{-k}\right\} .
\eea
Therefore, we obtain
\bea
\langle X(\theta)^{2} \rangle
=-\a'\frac{\ln r}{1-u\ln r}
-2\a'\sumk
\left(\frac{1}{k}\frac{u+(2k-u)r^{2k}}{k+u+(k-u)r^{2k}}\right),
\eea
after an appropriate subtraction of the divergent terms.
Then using the relation
\bea
\frac{d}{du}\ln Z_{1}^{\mb{tad}}
=-\tension \int^{2\pi}_{0}d\theta \langle X(\theta)^{2} \rangle,
\eea
we obtain
\bea
Z_{1}^{\mb{tad}}=&& 
\left\{
\frac{u e^{\gamma u}\Gamma(u)}
{\sqrt{1-u\ln r}}
\prod^{\infty}_{k=1}
\left(
\frac{1}{k}
\frac{k+u}{k+u+(k-u)r^{2k}}
\right)
\right\} \nonumber \\
&\times&(\mbox{ factors including}\:a,x_{0},r)
,
\eea
where $\gamma$ is Euler's constant.
In general, we obtain the partition function with
non-zero $u_{i}$ as
\bea
Z_{1}^{\mb{tad}}=&& \prod^{26}_{i=1}
\left\{
\frac{u_{i} e^{\gamma u_{i}}\Gamma(u_{i})}
{\sqrt{1-u_{i}\ln r}}
\prod^{\infty}_{k=1}
\left(
\frac{1}{k}
\frac{k+u_{i}}{k+u_{i}+(k-u_{i})r^{2k}}
\right)
\right\} \nonumber \\
&\times&(\mbox{ factors including}\:a,x_{0},r)
.
\eea
We can easily incorporate the $a$ dependent factor as
\bea
Z_{1}^{\mb{tad}}=&& e^{-a} 
\prod^{26}_{i=1}
\left\{
\frac{u_{i} e^{\gamma u_{i}}\Gamma(u_{i})}
{\sqrt{1-u_{i}\ln r}}
\prod^{\infty}_{k=1}
\left(
\frac{k+u_{i}}{k+u_{i}+(k-u_{i})r^{2k}}
\right)
\right\} \nonumber \\
&\times&(\mbox{ factors including}\:x_{0},r)
\nonumber \\
=&& Z_{0}(a,u_i)
\prod^{26}_{i=1}
\left\{
\sqrt{\frac{u_{i}}{1-u_{i}\ln r}}
\prod^{\infty}_{k=1}
\left(
\frac{k+u_{i}}{k+u_{i}+(k-u_{i})r^{2k}}
\right)
\right\} \nonumber \\
&\times&(\mbox{ factors including}\:x_{0},r)
,
\label{Z1ua}
\eea
where 
$Z_{0}(a,u_i)=e^{-a}\prod^{26}_{i=1}
\sqrt{u_{i}} e^{\gamma u_{i}}\Gamma(u_{i})$.
Here we have absorbed the factor 
$\prod^{26}_{i=1}\prod^{\infty}_{k=1}\frac{1}{k}$
into $e^{-a}$ and renormalized, as in Ref.~\citen{BSFT2}
.

\subsubsection{The $x_{0}^{\mu}$ dependent part}

Next, we fix the factor including $x_{0}^{\mu}$.
We determine the $x_{0}^{\mu}$ dependent factor as follows.
First, we map the annulus $M$ to a cylinder $M'$ with the diffeomorphism
\bea
z=e^{\frac{\rho}{t}},
\eea
where $\rho \equiv \xi^{1}+i\xi^{2}$ represents the 
complex coordinates on the cylinder.
The length of the cylinder is $\pi$ and 
its periodicity is $2\pi t$.
The inner circle of the annulus is mapped to one
side of the cylinder, where $\xi^{1}=-\pi$, and
the outer circle is mapped to the other side, where
$\xi^{1}=0$.
$t$ is related to $r$ as
\bea
r=e^{-\frac{\pi}{t}}.
\label{r}
\eea
Now $t$ is the moduli of the cylinder and the annulus,
which has already appeared in (\ref{total}).
Of course, the partition function does not change 
under this transformation, while the metric on the
cylinder is now
\bea
g_{\rho \bar{\rho}}=g_{\bar{\rho}\rho}
=\frac{|\exp(\frac{\rho}{t})|^{2}}{2t^{2}}.
\eea
Next, we carry out a Weyl transformation as
\bea
g_{\rho \bar{\rho}}
\rightarrow 
\frac{1}{|\exp(\frac{\rho}{t})|^{2}}
g_{\rho \bar{\rho}}
=\frac{1}{2t^{2}}.
\label{cyl-metric}
\eea
Note that this rescaling does not change the metric
where $\xi^{1}=0$, and therefore the boundary action is not
affected.
Furthermore, the bulk action possesses Weyl symmetry.
Thus the partition function does not change
under the above two successive transformations, even
though the boundary action breaks conformal invariance.
Now we have the worldsheet action on the cylinder $M'$
given by
\bea
S&=&\tension
\int_{M'}\! d\xi^{1} d\xi^{2} 
\partial_{a}X^{\mu}\partial^{a}X_{\mu} \nonumber \\
&+&
\tension \int_{\partial M'_{\mb{out}}}\!\!\!\! 
d\theta' \frac{1}{t}
\bigl\{2\a' a+\sum_{i=1}^{26}u_{i} X_{i}(\theta')^{2}
\bigr\}.
\label{cyl-action}
\eea

Next, we use the Minkowski signature on the cylinder and
calculate the Hamiltonian.
The mode expansion of $X^{\mu}$ on the cylinder is
\bea
X^{\mu}(\xi^{0},\xi^{1})
=x_{0}^{\mu}\frac{t-u_{\mu}\xi^{1}}{t+u_{\mu}\pi}
+(\mbox{oscillating modes}),
\eea
where we do not sum over $\mu$ on the right-hand side.
The worldsheet action contains only terms quadratic in the 
$X^{\mu}$ and a constant term.
Thus the zero-mode part of the Hamiltonian 
decouples from the oscillating-mode part.
The oscillating modes give the partition function
that is obtained in the case that $x_{0}^{\mu}=0$.
Therefore, we only have to calculate the
zero-mode part of the Hamiltonian, 
which yields the $x_{0}^{\mu}$ dependent factor.
We easily obtain
\bea
H&=&\tension 
\sum^{26}_{\mu=1}
\frac{u_{\mu}}{t+u_{\mu}\pi}
{\left(x_{0}^{\mu}\right)}^{2}
+
\tension
\frac{2\a' a}{t} \nonumber \\
&+&(\mbox{contribution of oscillating modes})
.
\eea
Therefore, the $x_{0}^{\mu}$ dependent factor of the
partition function can be obtained from
\bea
Z_{1}^{\mb{tad}}(x_{0}^{\mu})=
\mbox{Tr}\: e^{-H 2\pi t}
=e^{
-\frac{t}{2\a'}
\sum^{26}_{\mu=1}
\frac{u_{\mu}}{t+u_{\mu}\pi}
{\left(x_{0}^{\mu}\right)}^{2}
}
Z_{1}^{\mb{tad}}(x_{0}^{\mu}=0).
\eea
Note that this process also reproduces the $a$ dependent
factor explicitly.

Next, we integrate the $x_{0}^{\mu}$ dependent factor 
over $x_{0}^{\mu}$.
We regard the weight $w$ as a constant for simplicity, 
and we do not include it in the integrand.
This integration can be carried out easily, and we have
\bea
\int d^{26}\!\!x^{\mu}_{0}
\exp\Bigl(
-\frac{t}{2\a'}
\sum^{26}_{\mu=1}
\frac{u_{\mu}}{t+u_{\mu}\pi}
{\left(x_{0}^{\mu}\right)}^{2}
\Bigr)
=
\prod^{26}_{\mu=1}
\sqrt{
\frac{2\pi \a'}{t}
\frac{t+u_{\mu}\pi}{u_{\mu}}
}\:\:.
\label{x-factor}
\eea
We find that the inverse of the right-hand side of (\ref{x-factor})
is included in the right-hand side of (\ref{Z1ua}),
using the relation (\ref{r}).
Therefore, 
we obtain
\bea
\frac{1}{{(2\pi \a')}^{26/2}}
\int d^{26}\!\!x^{\mu}_{0} Z_{1}^{\mb{tad}}
=&&
Z_{0}
\left\{
\prod^{26}_{i=1}
\prod^{\infty}_{k=1}
\frac{k+u_{i}}{k+u_{i}+(k-u_{i})e^{-2\pi k/t}}
\right\} \nonumber \\
&\times& (\mbox{factors including } t)
.
\label{int-Z}
\eea
We note that finite values of the $u_i$ lead to the reduction
\bea
\sqrt{
\frac{t+u_{i}\pi}{tu_{i}}}
\sqrt{
\frac{u_i}{1-u_{i}\ln r}}
=1
\eea
in the calculation of (\ref{int-Z}).
This is one merit of using the
variables $u_i$ as regulators.

\subsubsection{The $t$ dependent part}

The $t$ dependent factor is obtained 
as follows. \cite{Suyama,Polchinski}
First we set $x_{0}^{\mu}=0$, and we treat only one of the $u_i$ 
as non-zero, again.
(Again we omit the subscript $i$.)
First, we calculate the energy-momentum tensor on the
cylinder $M'$. The energy-momentum tensor on the annulus $M$
is obtained as
\bea
\langle T_{zz} \rangle &=&
-\frac{1}{\a'}
\lim_{w\rightarrow z}
\left\{
\partial_{z}\partial_{w}G(z,w)
+\frac{\a'}{2}\frac{1}{(z-w)^{2}}
\right\} \nonumber \\
&=&
\frac{1}{z^{2}}
\left\{
\frac{1}{4}\frac{u}{1-u\ln r}
-\sumk
\frac{k(k-u)r^{2k}}{k+u+(k+u)r^{2k}}
\right\}.
\eea
The anti-holomorphic part is obtained similarly.
Then, the energy-momentum tensor on the cylinder $M'$
is
\bea
\langle T_{\rho\rho} \rangle &=&
\Bigl(\frac{dz}{d\rho}\Bigr)^{2}
\langle T_{zz} \rangle
+\frac{26}{12} \{z,\rho\} \nonumber \\
&=&
\frac{1}{t^{2}}
\left\{
\frac{1}{4}\frac{u}{1-u\ln r}
-\sumk
\frac{k(k-u)r^{2k}}{k+u+(k+u)r^{2k}}
-\frac{26}{24}
\right\}.
\eea
We use the following relation for the calculation
of the partition function;
\bea
\delta \ln Z_{1}^{tad}
=-\tension\int d^{2}\rho
\sqrt{g}
\left\{
\delta g_{\rho\rho}\langle T^{\rho\rho} \rangle
+
\delta g_{\bar{\rho}\bar{\rho}}
\langle T^{\bar{\rho}\bar{\rho}} \rangle
\right\}.
\eea
Suppose that we change the periodicity of the 
cylinder as $2\pi t \rightarrow 2\pi(t+\delta t)$,
while keeping the metric unchanged.
We can realize the same shift of the partition function
without changing the periodicity if we shift
the metric by the amount
\bea
\delta g_{\rho\rho}
=\delta g_{\bar{\rho}\bar{\rho}}
=-\frac{1}{2t}\frac{1}{t^{2}}\delta t,
\eea
instead.
Recall that the metric on the cylinder is given
in (\ref{cyl-metric}) as
\bea
g_{\rho\bar{\rho}}&=&g_{\bar{\rho}\rho}
=\frac{1}{2t^{2}},\nonumber \\
g_{\rho\rho}&=&g_{\bar{\rho}\bar{\rho}}
=0.
\eea
Using the above relations, we obtain the differential
equation
\bea
\frac{d}{dt} \ln Z_{1}^{\mb{tad}}
&=& 2\pi \langle T_{\rho\rho} \rangle \nonumber \\
&=& \frac{2\pi}{t^{2}}
\left\{
\frac{1}{4}\frac{u}{1-u\ln r}
-\sumk
\frac{k(k-u)r^{2k}}{k+u+(k+u)r^{2k}}
-\frac{26}{24}
\right\}.
\eea
Therefore, we obtain
\bea
Z_{1}^{\mb{tad}}=&&
e^{\frac{26\pi}{12t}}
\sqrt{\frac{t}{t+u\pi}}
\prod^{\infty}_{k=1}
\frac{1}{k+u+(k-u)e^{-2\pi k/t}} \nonumber \\
&\times&
(\mbox{factors including } a, u, x_{0}^{\mu}).
\eea
Note that the above procedure also exactly reproduces
the factors that include both $u$ and $t$.

\subsubsection{Conformal limit and the integration over 
the moduli $t$}

We obtain 
\bea
(2\pi \a')^{-26/2}
\int d^{26}\!\!x_{0}^{\mu}\: Z_{1}(a,u_i,x_{0}^{\mu},t)
=&&
Z_{0}(a,u_i)
e^{\frac{26\pi}{12t}}
\prod^{26}_{i=1}
\prod^{\infty}_{k=1}
\frac{k+u_i}{k+u_i+(k-u_i)e^{-2\pi k/t}} \nonumber \\
&&\times
\mbox{constant},
\label{anu-part}
\eea
after we incorporate the results of the previous subsections.
We absorb the ambiguity of the overall factor into 
the weight $w$.
Therefore, we set the constant factor to 1.

As the final step, we integrate (\ref{anu-part})
over the moduli $t$.
Of course, the gauge fixing procedure is justified
only in the conformal limit.
Therefore we take this limit and
perform the integration over $t$,
including the Faddeev-Popov determinant.

We take several types of conformal limits,
as mentioned in \S 3.
Using the result of BSFT, the disk partition function 
$Z_{0}(a,u_i)$ becomes
\bea
Z_{0}(a,u_i)
\longrightarrow
V_{n_0-1} T_{n_0-1}
\eea
in the conformal limit in which we take $n_0$ of the
quantities $u_i$
to be zero and the other $(26-n_0)$ quantities $u_i$ 
to be $\infty$.
Here $V_{n_0-1}$ is the volume of the D$(n_0-1)$-brane, and 
$T_{n_0-1}$ is the tension of the D$(n_0-1)$-brane 
at the tree level.

The Faddeev-Popov determinant on the annulus
is given in the usual way, since we do not have ghosts
in the boundary action.
It is obtained as\footnote{
For example, see Ref.~\citen{Polchinski}
.
}
\bea
\eta(it)^{2}
=\frac{1}{t}
e^{-\frac{2\pi}{12t}}
\prod^{\infty}_{k=1}
(1-e^{-2\pi k /t})^{2}.
\eea
Then, we obtain the correction term, which includes 
the contribution of one closed-string tadpole in the conformal
limit, as
\bea
g_{c}\int_{0}^{\infty} \frac{dt}{2t} 
\prod_{\mu} 
\int\!\! \frac{dx^{\mu}_0}{\sqrt{2\pi \a'}} 
w(x^{\mu}_0)Z_{1}^{\mb{tad}}(t,x^{\mu}_0,n_0)
=
g_{c}
V_{n_0-1} T_{n_0-1}
I^{\mb{tad}}(n_{0}),
\label{total-final}
\eea
where
\bea
I^{\mb{tad}}(n_{0})
&\equiv&
\int_{0}^{\infty} \frac{dt}{2t} 
\prod_{\mu} 
\int\!\! \frac{dx^{\mu}_0}{\sqrt{2\pi \a'}} 
w(x^{\mu}_0)
\frac{Z_{1}^{\mb{tad}}(t,x^{\mu}_0,n_0)}{Z_0(n_0)}
\nonumber \\
&=&
w \int^{\infty}_{0} \frac{dt}{2t^{2}}
e^{2\pi /t}
\prod^{\infty}_{k=1}
\left[
(1-e^{-2\pi k /t})^{n_{0}-24}
(1+e^{-2\pi k /t})^{-n_{0}}
\right].
\label{In0}
\eea
Note that $w$ is treated as a constant here.

\subsection{One-loop open-string partition function
on a D$(n_0-1)$-brane}

The ordinary one-loop partition functions for open-strings
are easily obtained, as described in string 
textbooks. \cite{Polchinski,GSW2}
We consider an annulus that is the same as the annulus $M$
in \S 4.1, except for the boundary conditions on the
inner circle.
The boundary conditions on the inner circle are the same
as those on the outer circle in this case.
The one-loop open-string partition function
on the D$(n_0-1)$-brane is given as
\bea
g_c
\int_0^{\infty} \frac{dt}{2t}
Z_1^{\mb{loop}}(t)
&=&
g_c V_{n_0-1} T_{n_0-1} I^{\mb{loop}}(n_0),
\eea
where
\bea
I^{\mb{loop}}(n_0)
\equiv
h
\int^{\infty}_{0} \frac{dt}{2t^{2}}
e^{2\pi /t}
t^{(26-n_0)/2}
\prod^{\infty}_{k=1}
(1-e^{-2\pi k /t})^{-24}.
\eea
Here $h$ is a constant that determines the ratio of 
the one-loop correction term to the leading term $Z_0$.
It is fixed by the stipulation that the theory be unitary,
as is usually the case in calculations of amplitudes.



\section{Calculation of the on-shell effective action}

\subsection{Conformal limit and the on-shell effective action}

We are now ready to discuss the on-shell effective action
of open-string tachyons on the background in which
closed-string tachyons are condensed.
We identify the vacua for open-string tachyons
as in the framework of BSFT as follows.\footnote{
We take the conformal limit while imposing the stationary condition
(\ref{statcond}). In \S 6 we discuss a problem that
can arise when we include $O(g_c)$ corrections.
}
Closed-string tachyons are assumed to already be condensed
here.

\vspace{0.5cm}

\begin{itemize}

\item {\bf perturbative vacuum}: $(a,u_i)\rightarrow (0,0)$\\
This corresponds to the situation in which open-string tachyons
are on a D25-brane.

\item {\bf non-perturbative vacuum}: $(a,u_i)\rightarrow (\infty,0)$\\
In this case, open-string tachyons are condensed completely.
We find that $Z_{0}=0$, and therefore $Z^{\mb{total}}=0$.

\item {\bf intermediate vacuum}: $a \rightarrow \infty$,
$n_{0}$ of the $u_i$ $\rightarrow 0$, and $n_{\infty}$ 
of the $u_i$
$\rightarrow \infty$ \\
We suppose that the open-string tachyons are on 
a D$(n_{0}-1)$-brane at this vacuum.
Open-string tachyons are partially condensed here.

\end{itemize}

\vspace{0.5cm}

\noindent

The on-shell effective action of open-string tachyons
at each vacuum is given in (\ref{total}).
Using the results of the previous section, we obtain
\bea
Z^{\mb{total}}=
V_{n_0-1}T_{n_0-1}
\left\{
1+g_c
\left(
I^{\mb{tad}}(n_0)+I^{\mb{loop}}(n_0)
\right)
+O(g_{c}^2)
\right\}.
\label{z^total}
\eea
We use an Euclidean signature in the spacetime metric in this
paper,
and thus $Z^{\mb{total}}/V_{n_0-1}$ is the energy density
of open-string tachyons.
If Sen's conjecture is still valid in the case that we
consider $O(g_{c})$ corrections, it corresponds to
the modified D$(n_0-1)$-brane tension.

\subsection{Elimination of the divergence in the effective action}

Next, we estimate the value of $Z^{\mb{total}}$ at each vacuum.
In this step, there is the difficulty that divergent terms exist.
The divergence of $I^{\mb{loop}}$ and $I^{\mb{tad}}$
in the small $t$ region corresponds to
the IR divergence in the closed-string channel, which
occurs when 
the inner circle of the annulus becomes infinitesimally small.
The divergence in the large $t$ region
is an IR divergence in the open-string channel,
due to the propagation of light open-string modes
circulating the annulus.

We now consider the divergence in the 
closed-string channel.
$I^{\mb{loop}}(n_0)$ can be rewritten as
\bea
I^{\mb{loop}}(n_{0})&=&
(2\pi)^{(26-n_0)/2}
h
\int^{\infty}_{0} \frac{ds}{4\pi}
e^{s}
s^{(n_0-26)/2}
\prod^{\infty}_{k=1}
(1-e^{-ks})^{-24}
\nonumber \\
&=&
(2\pi)^{(26-n_0)/2}
h
\int^{\infty}_{0} \frac{ds}{4\pi}
s^{(n_0-26)/2}
\left[
e^{s}+24+O(e^{-s})
\right],
\label{Iloop-div}
\eea
and $I^{\mb{tad}}(n_{0})$ 
can be written as
\bea
I^{\mb{tad}}(n_{0})&=&w
\int^{\infty}_{0} \frac{ds}{4\pi}
e^{s}
\prod^{\infty}_{k=1}
\left[
(1-e^{-ks})^{n_{0}-24}
(1+e^{-ks})^{-n_{0}}
\right] \nonumber \\
&=&w
\int^{\infty}_{0} \frac{ds}{4\pi}
\left[
e^{s}+(24-2n_0)+O(e^{-s})
\right],
\label{In0-div}
\eea
where $s=2\pi/t$.
The origin of the divergence in the closed-string channel
is light closed-string fields,
namely closed-string tachyons and dilatons.

\subsubsection{The divergence from tachyons}

The divergence coming from the first terms in the
last lines of (\ref{Iloop-div})
and (\ref{In0-div}) is due to zero-momentum
closed-string tachyons emitted as tadpoles into the
vacuum. 
This type of divergence is spurious.
We can obtain finite contributions from these terms 
by analytic continuation.
\cite{Callan-Coleman,Weinb-Wu,Weinb87,Marcus}
In this case, the regularized partition function can be a
complex number after the analytic continuation, in general.
Its imaginary part corresponds to the decay rate
of the unstable vacuum through the relation
\cite{Callan-Coleman,Weinb-Wu,Marcus}
\bea
\tau_{\mb{decay}}=-2 \mbox{Im}\{E\},
\eea
where $\tau_{\mb{decay}}$ is the decay rate of the unstable vacuum
per unit volume, and
$E$ is the regularized energy density of the unstable
vacuum. 
We assume $E=Z^{\mb{total}}/V_{n_0-1}$ here.
The imaginary part from closed-string tachyons
has to be removed, since we wish to
represent the effective action, which does not have
an instability due to closed-string tachyons.
We propose a method to cancel the imaginary part
in \S 5.3.

\subsubsection{The divergence from massless string fields}

A fatal divergence comes from the second terms in the
last lines of (\ref{Iloop-div})
and (\ref{In0-div}).
This divergence is due to on-shell dilatons emitted
into the vacuum.
It causes a conformal anomaly.
We have several methods to eliminate this divergence
for oriented strings.

\vspace{0.5cm}

\begin{itemize}

\item{Fischler-Susskind mechanism}

If we put appropriate vertex operators onto the
worldsheet, the background fields are shifted and
the conformal anomaly is absorbed.
This is called the
Fischler-Susskind mechanism. \cite{FS86-1,FS86-2}
The vertex operators to be inserted are those of
massless string fields, and the weights attached to
them are chosen so that the fatal divergence vanishes.
This mechanism causes the contribution
of zero-momentum tachyons to be non-zero in general.
This property is desirable in the treatment
of closed-string tachyon condensation.

\vspace{0.2cm}

\item{Cancellation between loops and tadpoles}

We have another method to cancel the
anomaly, which is presented in Ref.~\citen{Green88}.
Let us consider the case of $n_0=26$, for example.
If we choose the weight $w$ appropriately as
\bea
(24-2n_0)w=-(2\pi)^{(26-n_0)/2}24h,
\label{dil-cancel}
\eea
the dilaton tadpole coming from $I^{\mb{tad}}(26)$ and
that from $I^{\mb{loop}}(26)$ cancel.
However, we do not use this method in this paper
for the reason given in \S 5.3.

\end{itemize}

\vspace{0.5cm}

\subsubsection{The divergence in the open-string channel} 

We also have divergence from the open-string channel
in $I^{\mb{loop}}$ and $I^{\mb{tad}}$, in general.
This occurs when the radius of the inner circle becomes 
equal to that of the outer circle of the annulus.
We also have two types of divergence here,
one due to open-string tachyons and
the other to massless open-strings.

The divergence caused by open-string tachyons can also
be regularized using analytic continuation.
In general, an imaginary part appears after the
analytic continuation.
This reflects the instability
of the vacuum due to open-string tachyons.\footnote{
This treatment of the divergence from open-string tachyons, 
considering one-loop corrections to BSFT, is also
found in Refs.~\citen{BarKone} and \citen{Cra-Kra-Lar}.
}
The imaginary part is eliminated by open-string
tachyon condensation.
There is also a fatal divergence from massless open-string
fields here. This divergence can also be removed using the
Fischler-Susskind mechanism.\footnote{
See, for example, 
Refs.~\citen{Pol2}--~\citen{Lee-Ray}.
}

It might desirable to treat open-string tachyon condensation
using ``open-string tachyon tadpoles'', as we do in the
context of closed-string tachyon condensation.
However, our main purpose in this paper is investigation
of the effect of closed-string tachyon condensation.
For this reason, we do not consider the details of the mechanism
of open-string tachyon condensation here.
The effect of closed-string tachyon condensation on
open-string tachyon condensation is discussed
in \S 5.4.

\subsection{Construction of a consistent on-shell effective action}

Our goal in this paper is to describe physical
quantities on the stable vacuum of closed-strings.
Therefore $\tau_{\mb{decay}}$ coming from closed-string tachyons
should vanish, and the worldsheet should be conformally invariant.
Thus, natural criteria which 
on-shell effective actions should obey are as follows.

\vspace{0.5cm}

\begin{itemize}

\item {The conformal anomaly must be removed.}

\item {The imaginary part of the spacetime action
due to closed-string tachyons must vanish
after analytic continuation.}

\end{itemize}

\vspace{0.5cm}

These criteria lead us to the following procedure 
to obtain a consistent on-shell effective action.
The procedure we propose here consists of three steps.

\vspace{0.5cm}

\begin{enumerate}

\item{Analytic continuation to regularize tachyonic divergence:}

First, we make the spurious IR divergence from tachyons
finite using analytic continuation.

\vspace{0.2cm}

\item{Cancellation of the imaginary part:}

Next, we set the weight $w$ so that the
imaginary part in $Z^{\mb{total}}$
{\em due to closed-string tachyons} vanishes.
If we choose $w$ appropriately, the imaginary part
in open-string loops and that
in closed-string tachyon tadpoles cancel.

\vspace{0.2cm}

\item{Elimination of the conformal anomaly:}

In general, after the above two steps, 
we still have the fatal divergence from massless string modes.
In the last step, we use the Fischler-Susskind mechanism
to cancel the remaining divergence and make
the worldsheet conformally invariant.
In this step, we do not use cancellation of the
fatal divergences of loops and tadpoles mentioned
near (\ref{dil-cancel}) in \S 5.2.2, 
since we have already fixed the value
of $w$ in the previous step.
The condition for $w$ to cancel the
fatal divergence is not compatible with
the elimination of the imaginary part,
in general.

\end{enumerate}

\vspace{0.5cm}

In the second step, we remove the imaginary part from
closed-string tachyons.
That is, the appropriate choice of $w$ eliminates the
instability of the vacuum due to closed-string tachyons.
After the last step, strings exist in a curved spacetime,
in general. 
The insertion 
of closed-string tachyon tadpoles
and massless string-field vertex operators
into worldsheets effectively creates such a background.
The above procedure is the most natural method to obtain 
an on-shell spacetime action satisfying the above
criteria.
The strategy proposed here also acts as a constraint for
possible $w$.

It is important to note that there is still an ambiguity 
in the determination of the weight $w$.
To understand the above procedure,
let us estimate the imaginary part in the closed-string
channel in $I^{\mb{loop}}$.
We obtain
\bea
\mbox{Im}\{I^{\mb{loop}}(n_0)\}
&=&
\mbox{Im}
\bigl\{
\int_0^{\infty}\frac{ds}{4\pi}s^{(n_0-26)/2}e^{(1+i\epsilon)s}
\bigr\}
\nonumber \\
&=&
\frac{1}{4\pi}\frac{\pi}{\Gamma(\frac{26-n_0}{2})}
\eea
by using the relation
\bea
\mbox{Im}
\bigl\{
\int_0^{\infty}\frac{ds}{s}s^{-\alpha}e^{(\gamma+i\epsilon)s}
\bigr\}
=
\frac{\pi}{\Gamma(1+\alpha)}{\gamma}^{\alpha}.
\:\:\:\:(\mbox{for } \gamma > 0)
\eea
Therefore, we have an imaginary part in $I^{\mb{loop}}(n_0)$
in the case $n_0 \le 25$.
Contrastingly, $I^{\mb{tad}}(n_0)$ given by (\ref{In0-div}) 
does not have an imaginary part
from closed-string tachyons.
This is due to our treatment of $w$.
Note that we have regarded $w$ as a constant in the calculation
of $I^{\mb{tad}}(n_0)$ in \S 4.
However, $w$ can depend on $x^{\mu}_0$ in general.
If $w$ has a nontrivial dependence on  $x^{\mu}_0$,
$I^{\mb{tad}}(n_0)$ also can be a complex number after the
analytic continuation.
For example, let us assume that $w$ is given by
\bea
w(x_0)=w_{n_0}(2\pi \a')^{(26-n_0)/2}\:\delta^{(26-n_0)}(x_0^j-0),
\label{wn0-1}
\eea
where $x_0^j$ is the spacetime coordinate perpendicular
to the D$(n_0-1)$-brane.
Then $I^{\mb{tad}}(n_0)$ is obtained as
\bea
I^{\mb{tad}}(n_0)&=&
\int_{0}^{\infty} \frac{dt}{2t} 
\prod_{\mu} 
\int\!\! \frac{dx^{\mu}_0}{\sqrt{2\pi \a'}} 
w(x^{\mu}_0)
\frac{Z_{1}^{\mb{tad}}(t,x^{\mu}_0,n_0)}{Z_0(n_0)}
\nonumber \\
&=&
w_{n_0}
\int^{\infty}_{0} \frac{dt}{2t^{2}}
e^{2\pi /t}
{\left(
\frac{t}{\pi}
\right)}^{\frac{26-n_0}{2}}
\prod^{\infty}_{k=1}
\left[
(1-e^{-2\pi k /t})^{n_{0}-24}
(1+e^{-2\pi k /t})^{-n_{0}}
\right]
\nonumber \\
&=&
2^{(26-n_0)/2} w_{n_0}
\int^{\infty}_{0} \frac{ds}{4\pi}
e^{s}
s^{(n_0-26)/2}
\prod^{\infty}_{k=1}
\left[
(1-e^{-ks})^{n_{0}-24}
(1+e^{-ks})^{-n_{0}}
\right] \nonumber \\
&=&
2^{(26-n_0)/2} w_{n_0}
\int^{\infty}_{0} \frac{ds}{4\pi}
s^{(n_0-26)/2}
\left[
e^{s}+(24-2n_0)+O(e^{-s})
\right].
\eea
In this case, if we set $w_{n_0}$ as
\bea
w_{n_0}=-\pi^{(26-n_0)/2}h,
\label{wn0-2}
\eea
the imaginary part in the closed-string channel
in $I^{\mb{tad}}(n_0)$ cancels that in $I^{\mb{loop}}(n_0)$.

After the elimination of the imaginary part using
the relation (\ref{wn0-1}) and (\ref{wn0-2}), we obtain
\bea
I^{\mb{loop}}(n_0)+I^{\mb{tad}}(n_0)
=
{2\pi}^{(26-n_0)/2} h
\int^{\infty}_{0} \!\! \frac{ds}{4\pi}
s^{(n_0-26)/2}
\left[
2n_0+O(e^{-s})
\right].
\eea
We still have a fatal divergence in the first term in 
the integral if $n_0 \ne 0$.
This divergence can be renormalized into the background
fields using the Fischler-Susskind mechanism.

The conditions (\ref{wn0-1}) and (\ref{wn0-2}) give us
one of natural choice for $w$.
However, this is only one of the possible choices, and
it is not unique.
We still have an ambiguity in determining the weight $w$.
The nontrivial dependence of $w$ on $x^{\mu}_0$ gives
a nontrivial relation between $w$ and $h$ instead 
of (\ref{wn0-2}).
Of course, the weight $w$, which is assumed to correspond to
the expectation value of closed-string tachyons,
should be determined dynamically.
However, it is very difficult to treat closed-string tachyons
dynamically at this stage.
The rigorous determination of $w$ remains to be studied.

\subsection{The effect on open-string tachyon condensation}

The effect of closed-string tachyon condensation
on open-string tachyon condensation is also interesting
to study.
Although we cannot give rigorous discussion, because
there is an ambiguity in the manner of determining the weight $w$,
we give preliminary discussion on this
topic.

In general, $Z_1^{\mb{tad}}$ can
affect open-string tachyon condensation.
For example, the choice of $w$ as (\ref{wn0-1}) and (\ref{wn0-2})
creates an interesting situation.
In this case, we have
\bea
I^{\mb{loop}}(n_0)+I^{\mb{tad}}(n_0)
&=&
h
\int^{\infty}_{0} \frac{dt}{2t^{2}}
e^{2\pi /t}
t^{(26-n_{0})/2}
\prod^{\infty}_{k=1}
\left[
(1-e^{-2\pi k /t})^{n_{0}-24}
(1+e^{-2\pi k /t})^{-n_{0}}
\right]
\nonumber \\
&-&
h
\int^{\infty}_{0} \frac{dt}{2t^{2}}
e^{2\pi /t}
t^{(26-n_{0})/2}
\prod^{\infty}_{k=1}
\left[
(1-e^{-2\pi k /t})^{-24}
\right].
\eea
The divergence in the open-string channel occurs
in the large $t$ region, in general.
However, if $n_{0}=0$, $I^{\mb{loop}}(n_0)+I^{\mb{tad}}(n_0)$
vanishes completely.
Thus, the ``D$(-1)$-brane'' has no instability
due to open-string tachyons at order $g_c$
in this case.
Of course, we {\em cannot} conclude immediately that 
bosonic D$(-1)$-branes
are stable on the nontrivial background, which the procedure in \S 5.3
with the choice of the weight as (\ref{wn0-1}) and (\ref{wn0-2})
creates.
However, the above considerations seem to imply the possibility
of the existence of some nontrivial stable vacuum
for open bosonic strings.
Further study is needed to obtain any definite
conclusion in this regard.
The any case, we can calculate the modified energy density 
$Z^{\mb{total}}/V_{n_0-1}$
of open-string tachyons as (\ref{z^total}) up to 
$O(g_c^2)$, using the above procedure.
If Sen's conjecture still holds for the case in which we include
$O(g_c)$ corrections, the energy density $Z^{\mb{total}}/V_{n_0-1}$
corresponds to the corrected tension of the D$(n_0-1)$-brane.

\section{Conclusion and discussion}

We have proposed a basic strategy to obtain a modified 
on-shell effective action for open-string tachyons, 
which includes loop corrections and the effect of closed-string
tachyon condensation.
The basic tool we used is the sigma-model approach.
We included a boundary term in the worldsheet action
to utilize the results of BSFT in defining the on-shell
partition function.
To incorporate the effect of closed-string tachyon condensation,
we inserted D-instanton-like macroscopic holes
into the worldsheets, and we regarded them as
closed-string tachyon tadpoles.
We presented a
natural procedure to obtain a consistent effective
action, using the tadpoles.
The procedure we proposed provides a constraint for the
weight that is attached to the tadpole, 
although we cannot determine the weight  rigorously
at this stage.
The instability due to closed-string tachyons vanishes
through this procedure.
We obtain a finite energy density of open-string tachyons
with this method, in principle.
The most important problem is determining how to derive 
the weight $w$ rigorously.
At present, this involves an ambiguity.
The rigorous determination of $w$ is left as an important
future project.

Now we give some remarks.
First,
the sigma-model approach using the results of BSFT presented
in this paper can be used to estimate the energy density
of open-string tachyons even if we do not consider the
effect of closed-string tachyons, in principle,
as $Z^{\mb{total}}$ with $w=0$ represents the
on-shell effective action of
open-string tachyons with one-loop quantum 
corrections.\footnote{
In this case, we should consider, for example, 
the Fischler-Susskind
mechanism or $SO(2^{13})$ unoriented strings to eliminate 
the conformal anomaly from loop diagrams.
}
Therefore, the method proposed here should be useful
to study the quantum corrections
to the D-brane tension if we regard $Z^{\mb{total}}/V_{n_0-1}$ 
as the tension of a D$(n_0-1)$-brane.\footnote{
A calculation of quantum corrections to the tension of
D-branes is found in Ref.~\citen{BCFPM}
.
}
However, there is an unresolved point.
We took the conformal limit while imposing the stationary condition
(\ref{statcond}) in the derivation of the on-shell effective
action of open-string tachyons.
The stationary condition (\ref{statcond}) is 
the equation of motion for the variable $a$ in 
the framework of BSFT, as mentioned in the Appendix.
In this sense, when we include $O(g_c)$
corrections in the effective action,
the condition (\ref{statcond}) can
possess a term
\bea
a=\sum^{26}_{i=1}
\left(
-u_i+ u_i\frac{\p}{\p u_i}\ln Z_0(u_i)
\right)
+
\sum_{s}
\left(
\frac{1}{2}-\frac{u_s x_s^2}{2\alpha'} 
\right)
+g_c f(u_i),
\label{statcond-gc}
\eea
as a correction, where $f(u_i)$ is a function of 
the $u_i$ that we cannot
determine rigorously at this stage. 
If $f(u_i)$ does not vanish in the conformal
limit, we have an extra factor $e^{-g_c f}$ in the
on-shell effective action, and then we have an additional 
$O(g_c)$ correction
term $-g_c f Z_0$ in the effective action.
So far, we have ignored the possibility of the existence of
the above correction.
We cannot make a rigorous argument to treat this problem,
since we do not have an extended BSFT including
$O(g_c)$ corrections, from which we should derive the
explicit form of $f(u_i)$. 
Thus there is room for further investigation of this problem.
In this article, we simply assume that $f(u_i)$ is not divergent
and that it does not affect the procedure for
obtaining a consistent on-shell effective action presented in
\S 5, even if $f(u_i)$ is non-zero and finite
in the conformal limit.

Second,
the most difficult problem regarding the topic of this
paper is the rigorous
determination of the weight $w(x_0^{\mu})$.
It would be interesting to attempt utilizing string
dualities to obtain some useful information concerning the weight. 

Third,
the choice of the weight in (\ref{wn0-1}) and (\ref{wn0-2}) means that
the D-instanton-like tadpoles are localized on the
D$(n_0-1)$-brane.
Inhomogeneous distributions of closed-string
tadpoles like this also suggest the modification of the spacetime
structure from flat spacetime.
It is pointed out that the spacetime positions where we can
fix ends of strings using Dirichlet
boundary conditions 
are restricted in the subspace of the total target space
if dilatons have a nontrivial dependence on the spacetime
coordinates. \cite{Nakamura}
We have to check the consistency of the Dirichlet boundary
conditions we used
with the configuration of the background dilatons
created by the Fischler-Susskind mechanism.
We leave this problem for further investigation.
To study the relation between closed-string 
tachyon condensation and the dynamical nature of 
spacetime structure is a very interesting and an important
project.

Fourth,
there is also a fundamental question concerning the value of $w$.
We implicitly assumed that $g_c w$ is sufficiently small, so that
the expansion with respect to the topology of worldsheets
gives us a good approximation.
However, the expectation value of fields in
point-particle field theories can be of the order of 
$($coupling$)^{-1}$ in general.
Actually, all the diagrams with tadpoles without loops
are at the tree level, for any number of
the attached tadpoles, in point-particle field theories.
In this sense, $g_c w$ also can be of order 1,
and we might have to consider all the
worldsheets that include multiple tadpoles,
as well as the annulus $M$ we consider in this paper.
This problem will become clear if we obtain the correct
value of $w$.
In any case, the basic strategy we presented in this paper
to obtain a consistent effective action is still
natural, though the cancellation of the imaginary
parts only within the annulus diagrams, done in
\S 5, is not correct in this case.

Finally, 
we treated oriented strings in this paper for simplicity.
If we consider unoriented strings, we have another
method to eliminate the fatal divergence;
we can cancel this divergence if we include cross-cap
diagrams and make an appropriate choice of the gauge group.
In this case, the situation seems to be more complicated.
We may be able to make the effective action finite and
remove the instability due to closed-string tachyons
using a combination of several methods,
the Fischler-Susskind mechanism,
the inclusion of cross caps with the appropriate choice 
of gauge group, and a suitable choice of the
weight of closed-string tachyon tadpoles.
The required gauge group could be different from
the ordinary $SO(2^{13})$ group, in general.
Consideration of the spacetime gauge anomaly is also
important to make the model consistent.
Construction of consistent (anomaly free) string theories
using D-instanton-like objects with appropriate weights,
as first pointed out by Green in Ref.~\citen{Green88},
may be interesting,
even without considering the problem of tachyons.

\section*{Acknowledgements}
The author would like to thank H.\ Kawai and S.\ Moriyama for valuable 
discussions and comments.

\appendix
\section{A Comment on the Stationary Conditions in BSFT} 

There are some delicate problems in taking the conformal limit
in BSFT.
We discuss them in detail here.
Let us consider the stationary condition for the
variable $a$. 
The condition $\frac{\p}{\p a}S(a,u_i)=0$ for the action
given by (\ref{BSFT}) and (\ref{Z-zero}) yields this
condition as
\bea
a&=&\sum^{26}_{i=1}
\left(
-u_i+ u_i\frac{\p}{\p u_i}\ln Z_0(u_i)
\right).
\label{statcond-naive}
\eea
However, we find that
(\ref{statcond-naive}) gives
\bea
\lim_{u_i \rightarrow 0}a
&=&-\frac{26}{2} \ne 0
\eea
if we take the limit $u_i \rightarrow 0$.
This implies that the equation of motion for $a$,
$\frac{\p}{\p a}S(a,u_i)=0$, is not compatible with the
conformally invariant solution 
$(a_{\ast},u_{\ast i})=(0,0)$.
This is not a desirable situation.

We find the same problem in another place.
For small values of $u_i$ we obtain
\bea
S(a,u_i)&=&
A
(a+1+\frac{26}{2})e^{-a}
\prod^{26}_{i=1}\frac{1}{\sqrt{u_i}} \nonumber \\
&+&
A
e^{-a}\bigl(\sum^{26}_{i=1} u_i \bigr)
\prod^{26}_{i=1}\frac{1}{\sqrt{u_i}} \nonumber \\
&+&\cdots,
\label{S-u-small}
\eea
where the first term corresponds to the potential term
and the second term corresponds to the kinetic term
in the derivative truncated action $S(T_{\mb{open}})$
for open-string tachyons $T_{\mb{open}}$. \cite{GerSha,KMM1}
$S(T_{\mb{open}})$ is given by \cite{GerSha,KMM1}
\bea
S(T_{\mb{open}})=
A
\int\!\! \frac{d^{26}x}{{\sqrt{2\pi \alpha'}}^{26}}
\left\{
\alpha'
e^{-T_{\mbt{open}}}(\p T_{\mb{open}})^{2}
+
e^{-T_{\mbt{open}}}(1+T_{\mb{open}})
\right\},
\label{ST}
\eea
where we have defined $T_{\mb{open}}$ as
$T_{\mb{open}}=a+\sum_i^{26} \frac{u_i x^{2}_i}{2\alpha'}$,
with $x^i$ the zero mode of $X^i$.
Here we find from (\ref{ST}) that 
the factor $A$ should be fixed as 
$A=T_{25} (2\pi\alpha')^{13}$.
The potential term in (\ref{ST}) becomes
\bea
A
\int\!\! \frac{d^{26}x}{{\sqrt{2\pi \alpha'}}^{26}}
e^{-a}(1+a)
\label{correct-pot}
\eea
in the small $u_i$ region, while (\ref{S-u-small})
implies
\bea
\mbox{potential} \propto 
e^{-a}\bigl(a+1+\frac{26}{2}\bigr).
\label{wrong-pot}
\eea
Here we again find the additional term $\frac{26}{2}$
in (\ref{wrong-pot}).

The origin of the additional term is found in the 
following calculation.
If we calculate the potential term in (\ref{ST}) explicitly,
we obtain
\bea
A
\int\!\! \frac{d^{26}x}{{\sqrt{2\pi \alpha'}}^{26}}
e^{-T_{\mbt{open}}}(1+T_{\mb{open}})
&=&
A
e^{-a}(a+1)
\int\!\! \frac{d^{26}x}{{\sqrt{2\pi \alpha'}}^{26}}
\exp\bigl\{-\sum_{i=1}^{26} \frac{u_i x_i^2}{2\alpha'}\bigr\}
\nonumber \\
+
&A&
e^{-a}\!\! 
\int\!\! \frac{d^{26}x}{{\sqrt{2\pi \alpha'}}^{26}}
\frac{u_i x_i^2}{2\alpha'} 
\exp\bigl\{
-\sum_{i=1}^{26} \frac{u_i x_i^2}{2\alpha'}
\bigr\}.
\label{pot-int}
\eea
The first term on the right-hand side gives (\ref{correct-pot}).
We emphasize that
the second term needs delicate treatment when we
take the conformal limit $u_i \rightarrow 0$.
If we take the limit {\em before} we perform the integration
over $x^i$, the second term in (\ref{pot-int}) vanishes,
while if we take the limit {\em after} the integration,
it does not vanish, and
it yields an additional term.

We have learned from the above discussion
that we should take the conformal limit $u_i \rightarrow 0$
{\em before} we perform the integration over $x^i$.
However, the BSFT action $S(a,u_i)$ does not possess the integral
explicitly; this integration has already been
performed as an integration over
the zero modes of the $X^i$ when we perform the path integral
in the derivation of the partition function $Z_0$.
Thus, we apply the following trick.
First, we insert the identity
\bea
1&=&\int \frac{d^{26}y}{{\sqrt{\pi}}^{26}}
\exp\bigl\{-\sum_{i=1}^{26} u_i y_i^2\bigr\}
\nonumber \\
&=&\int \frac{d^{26}x}{{\sqrt{b\pi \alpha'}}^{26}}
\exp\bigl\{-\sum_{i=1}^{26} \frac{u_i x_i^2}{b\alpha'}\bigr\},
\:\:\:\:
(\mbox{for } u_i>0 )
\eea
into the partition function $Z_0$, where
$y^i$ is a dimensionless parameter.
We have also introduced the parameter $b$ as
\bea
y^{i}=\frac{x^{i}}{\sqrt{b\alpha'}}.
\label{scale}
\eea
Now, we wish to regard $x^{i}$ in (\ref{scale})
as the zero mode of $X^{i}$, which corresponds to
the spacetime coordinates.
In the partition function $Z_0$, we have the integral
over the zero mode as
\bea
\int d^{26}x
\exp\bigl\{-a-\sum_{i=1}^{26} \frac{u_i x_i^2}{2\alpha'}\bigr\},
\eea
which is extracted from (\ref{Z-lambda}) and
the boundary term in (\ref{worldsheet}).
Therefore we naturally set $b=2$, and we insert
\bea
1&=&\int \frac{d^{26}x}{{\sqrt{2\pi\alpha'}}^{26}}
\exp\bigl\{-\sum_{i=1}^{26} \frac{u_i x_i^2}{2\alpha'}\bigr\}
\eea
into $Z_0$.
Then the action $S(a,u_i)$ is rewritten as
\bea
S(a,u_i)
=
\int \frac{d^{26}x}{{\sqrt{2\pi\alpha'}}^{26}}
{\cal L}(a,u_i,x^i),
\eea
where
\bea
{\cal L}(a,u_i,x^i)
&=&
{\cal Z}_0^x 
\bigl(
1+a+\sum_i u_i -\sum_i u_i \frac{\p}{\p u_i}\ln {\cal Z}_0^x 
\bigr),
\nonumber \\
{\cal Z}_0^x(a,u_i,x^i)
&=&
A e^{-a}
\prod_{i=1}^{26}
u_i e^{\gamma u_i}\Gamma(u_i) e^{-\frac{u_i x_i^2}{2\alpha'}}.
\eea
Taking the conformal limit $u_i \rightarrow 0$ before
the integration over the spacetime coordinates means
that we should take the limit while enforcing the condition
$\frac{\p}{\p a} {\cal L}=0$, not the condition
$\frac{\p}{\p a} S=0$.
Then we obtain the condition
\bea
a
&=&
\sum^{26}_{i=1}
\left(
-u_i+ u_i\frac{\p}{\p u_i}\ln {\cal Z}_0^x 
\right)
\nonumber \\
&=&
\sum^{26}_{i=1}
\left(
-u_i+ u_i\frac{\p}{\p u_i}\ln Z_0
+\frac{1}{2}-\frac{u_i x_i^2}{2\alpha'} 
\right),
\label{stat-cond-small-u}
\eea
instead of (\ref{statcond-naive}), from $\frac{\p}{\p a} {\cal L}=0$.
The condition (\ref{stat-cond-small-u}) gives 
$a \rightarrow 0$ when we take
$u_i \rightarrow 0$.
We also find
\bea
\lim_{u_i \rightarrow 0}{\cal L}=(1+a)e^{-a}
\eea
if we take the limit subject to (\ref{stat-cond-small-u}).
Therefore, our problem is solved.

However, we encounter the opposite situation when
we take the limit $u_i \rightarrow \infty$.
In this case
we find that we should take the limit $u_i \rightarrow \infty$
{\em after} the integration over the spacetime coordinates.
For example,
if we take $u_1 \rightarrow \infty$ while maintaining the condition 
(\ref{stat-cond-small-u}), we find that ${\cal L}$ diverges as
${\cal L}\rightarrow O(\sqrt{u_1})$.
On the other hand, we obtain the correct result
\bea
\lim_{u_1 \rightarrow \infty} S(a,u_i)
=\frac{2\pi\sqrt{\alpha'}}{\mbox{Vol}}S(0,0),
\eea
where Vol is the spacetime volume in the $x^1$ direction,
if we take
the limit $u_1 \rightarrow \infty$ 
{\em after} the integration
over the spacetime coordinate $x^1$.
Thus, we obtain the correct ratio of
tensions of D-branes if
we take the limit while enforcing the condition
\bea
a=-u_1+ u_1\frac{\p}{\p u_1}\ln Z_0,
\label{KMM-cond}
\eea
which corresponds to (\ref{statcond-naive}).
In this case, we take $u_i \rightarrow 0$ subject to
(\ref{stat-cond-small-u}) for $i \ne 1$.
The condition (\ref{KMM-cond}) is used to derive
the ratio of tensions of D-branes in Ref.~\citen{KMM1}.

We thus find that the most suitable stationary condition is
\bea
a=
\sum^{26}_{i=1}
\left(
-u_i+ u_i\frac{\p}{\p u_i}\ln Z_0
\right)
+
\sum_{s}
\left(
\frac{1}{2}-\frac{u_s x_s^2}{2\alpha'} 
\right),
\label{correct-stat}
\eea
where the $u_s$ are the variables that are taken to
zero in the conformal limit.
That is, we have Neumann boundary conditions in the $x^s$ direction
after we take the conformal limit
and the number of coordinates $x^s$ is $n_0$.
This is the condition given in (\ref{statcond}).
In this case, $S(a,u_i)$ should be written
\bea
S(a,u_i)
&=&
\bigl(
\prod_s
\int \frac{d x^s}{\sqrt{2\pi\alpha'}}
\bigr)
{\cal Z}_0 
\bigl(
1+a+\sum_i u_i -\sum_i u_i \frac{\p}{\p u_i}\ln {\cal Z}_0 
\bigr),
\nonumber \\
{\cal Z}_0(a,u_i,x^s)
&=&
A e^{-a}
\bigl(
\prod_{i=1}^{26}
\sqrt{u_i} e^{\gamma u_i}\Gamma(u_i) 
\bigr)
\prod_s
\sqrt{u_s}
e^{-\frac{u_s x_s^2}{2\alpha'}}.
\label{rewrite-S}
\eea
Note that only the integration over $x^s$ has to be restored.
We can easily check that $S(a,u_i)=Z(a,u_i)$ if
the stationary condition (\ref{correct-stat})
is satisfied.
Of course, the variable $a$ should be inside the
integral over $x^s$ when we use the condition
(\ref{correct-stat}).
Thus we should take the conformal limit while maintaining 
(\ref{correct-stat}), which is also given
in (\ref{statcond}), and we have to rewrite
$S(a,u_i)=Z(a,u_i)$ in the form of (\ref{rewrite-S}).
The zero-mode integrals in the $x^s$ directions should
be performed after we take the limit
$u_s \rightarrow 0$.
We can obtain the correct value for the on-shell effective 
action of open-string tachyons if we use the stationary
condition (\ref{correct-stat}).

We now give a final comment.
The second term in (\ref{pot-int}) always vanishes if we
consider BSFT in a finite volume spacetime.
In this case, we can use the condition (\ref{statcond-naive}).
Therefore, considering a finite volume spacetime as an IR
regularization and using the condition (\ref{statcond-naive})
represents another solution.

\end{document}